# An ISOPHOT study* of the disk of galaxy NGC 6946: 60 μm infrared and radio continuum correlation

N. Y. Lu[1], G. Helou[1], R. Tuffs[2], C. Xu[2], S. Malhotra[1], M. W. Werner[3], and H. Thronson[4]

[1] Infrared Processing and Analysis Center, MS 100-22, California Institute of Technology, Pasadena, CA 91125, U.S.A.
[2] Max-Planck Institut für Kernphysik, Saupfercheckweg 1, D-69117 Heidelberg, Germany
[3] Jet propulsion Laboratory, MS 233-303, California Institute of Technology, Pasadena, CA 91109, U.S.A.
[4] Wyoming Infrared Obs., Univ. of Wyoming, Laramie, WY 82071, and Code SR, NASA HQs, Washington DC 20546, U.S.A.

**Abstract.** We use the ISOPHOT 60 μm image of Tuffs et al. (1996) to study the 60 μm light distribution and its correlation with radio continuum within the disk of spiral galaxy NGC 6946, and to resolve the long-standing IRAS controversy as to whether the infrared-to-radio continuum ratio, $Q$, decreases radially out in galaxy disks. Our main results are: (1) The disk at 60 μm also appears to follow an exponential fall-off. Its e-folding scale-length lies between $1'.3$ and an upper limit of $1'.9$, obtained from the mean radial profile between $1'.5$ and $4'$ in radius. This scale length is smaller than that derived from the radio continuum data at either 20 cm or 6 cm, with or without the thermal component removed, at a significance level of $\gtrsim 2\sigma$. (2) From $1'.5$ to $4'$ in radius, the 60 μm-to-radio continuum surface brightness ratio, $Q$, decreases by a factor of 1.5 to 1.7 on average. These quantitative results agree with that of Marsh & Helou (1995) using IRAS HiRes data, but not with that of Fitt et al. (1992) based on IRAS CPC data. Our results also fit into the picture of the radio disk as a "smeared" version of the infrared disk, but are not consistent with the picture in which the variation of $Q$ is a result of the thermal and non-thermal radio components having two distinct, constant values of the infrared-to-radio ratio.

## 1. Introduction

IRAS has revealed a good correlation between the thermal far-infrared (FIR) emission and the primarily non-thermal radio continuum emission for spiral galaxies (e.g., Dickey & Salpeter 1984). More detailed studies of a dozen or so nearby galaxies using IRAS HiRes images showed that the FIR-to-radio surface brightness (SB) ratio, $Q$, generally decreases with increasing radius in these galaxies (e.g., Marsh & Helou 1995). This systematic variation of $Q$ may play a crucial role in understanding the physics behind the global infrared-radio correlation among galaxies (Helou & Bicay 1993). However, using the FIR data from IRAS Chopped Photometric Channel (CPC) instrument on a smaller set of galaxies, Fitt et al. (1992) showed little evidence for this systematic behavior of $Q$. Since the radio data used in these studies were all drawn from the same source, these conflicting results obviously have their root in the IRAS data used. The Infrared Space Observatory (ISO; Kessler et al. 1996) offers a unique opportunity to resolve this long-standing controversy by providing independent FIR measurements as we show here on NGC 6946, a spiral galaxy common to both of the above studies.

Given its large angular size and nearly face-on appearance, NGC 6946 is among the best objects for studying the FIR light distribution (e.g., Engargiola 1991) and its correlation with radio continuum within the galaxy disk. While the SB distributions at various FIR wavelengths are discussed mainly in Tuffs et al. (1996), we limit this letter to a comparative study of the 60 μm and radio continuum emissions within the galaxy disk. We briefly describe the data in §2, discuss the infrared and radio light distributions in §3, and study the spatial variation of $Q$ in §4.

## 2. Infrared and Radio Images

### 2.1. ISOPHOT Image at 60 μm

The 60 μm image used here was obtained with the ISOPHOT (hereafter PHT; Lemke et al. 1996) and reported in Tuffs et al. (1996), where the observational and data reduction details are also given. The r.m.s. noise and

*Send offprint requests to*: N. Y. Lu at lu@ipac.caltech.edu
* Based on observations with ISO, an ESA project with instruments funded by ESA Member States (especially the PI countries: France, Germany, the Netherlands and the United Kingdom) and with the participation of ISAS and NASA.

effective angular resolution of the image are about 0.7 MJy sr$^{-1}$ and 52″ ($\approx$ 1.4 kpc at the distance of 5.5 Mpc of Pierce 1994), respectively.

Two potential systematic errors on the disk SB measurement are the stray-light contamination from the bright galaxy nucleus and the detector transient effect. A preliminary analysis indicated that the stray-light contamination is unlikely to be significant at outer radii and that the detector transient effect is probably significant only within one resolution element of the galaxy nucleus (see Tuffs et al. 1996). Nevertheless, we derive below a *conservative* upper limit on all possible systematic errors by a comparison with IRAS data, and roughly estimate the transient effect on the disk SB in §3.

**Table 1.** Comparison of the ISOPHOT and IRAS images

| $r_0$ (arcmin) | < 1 | 1-2 | 2-3 | 3-4 | 4-4.5 | 4.5-5 |
|---|---|---|---|---|---|---|
| No. of cells | 7 | 18 | 32 | 42 | 24 | 30 |
| $<P>$ | 80.0 | 39.5 | 27.2 | 14.9 | 8.9 | 6.7 |
| $<P>/<I>$ | 0.4 | 0.8 | 1.0 | 0.9 | 1.2 | 1.4 |
| r.m.s.$(P-I)$ | 71.9 | 19.5 | 8.3 | 6.3 | 4.2 | 3.1 |

The IRAS 60 $\mu$m image used in the comparison was constructed using the HiRes software with 20 iteration steps, giving an effective resolution of 40″ to 60″ and an r.m.s. noise of 0.66 MJy sr$^{-1}$. The result of the comparison is summarized in Table 1 which gives, for each radial bin in the face-on radius $r_0$, the number of cells (to be explained below); the mean PHT SB, $<P>$, in MJy sr$^{-1}$; the ratio of $<P>$ to the mean IRAS SB $<I>$; and the r.m.s. of $(P-I)$ in MJy sr$^{-1}$. In evaluating these parameters, each of the galaxy images was divided into cells of 40″ × 40″. The radius, $r_0$, of each cell in the plane of the galaxy was determined using an inclination of 30° and a position angle of 69° (N to E). Except for $r_0 \lesssim 1'$ where the PHT data are known to underestimate the true SB (Tuffs et al. 1996), the two maps agree with each other on average in the inner disk. The PHT map however is systematically brighter than the IRAS map at outer radii. In the absence of definitive evidence favoring either map, this table can serve as an upper-limit indicator on systematic errors affecting the PHT map at a given radius.

### 2.2. Radio Continuum Images

The radio continuum images at 4.8 GHz ($\sim$ 6 cm) and 1.465 GHz ($\sim$ 20 cm) were taken from Beck & Hoernes (1996) and Beck (1991), respectively. The 6 cm radio image combined a low-resolution map with a high-resolution VLA D-array map which has a beam size of 12″.5, resulting in a more accurate measurement of the total radio power. The VLA 20 cm radio image has a beam size of 42″. We have processed both the images to be comparable to the PHT image in resolution and pixel size. The r.m.s. noises of the final radio images are on the order of 10$^{-3}$ MJy sr$^{-1}$ each.

Using these radio maps, we constructed a radial distribution of the mean radio spectral index $\alpha$, defined as in $S_\nu \propto \nu^{-\alpha}$. The result is very similar to that of Klein et al. (1982; see their Fig. 5) for $r_0 < 4'$, with a peak spectral index of 0.78 ± 0.05 near $r_0 = 1'.5$. Beyond $r_0 = 4'$, however, $\alpha$ starts to decrease with increasing radius. Lacking a clear physical explanation and being inconsistent with other studies (van der Kruit et al. 1977; Klein et al. 1982), this $\alpha$ behavior at large radii probably arises from the underrepresentation of the SB at 20 cm relative to that at 6 cm, especially at the western side of the galaxy disk. As a result, we shall limit our quantitative FIR/radio comparison to $r_0 < 4'$. In order to separate the thermal component from the non-thermal component, we assumed, as in Klein et al. (1982), constant spectral indices of 0.1 and 0.8 for these two components, respectively. Outside the nucleus but within $r_0 = 4'$, the thermal component accounts for about 18% (11%) of the total radio emission at 6 cm (20 cm) on average.

## 3. FIR and Radio-Continuum Light Distributions

As illustrated in Fig. 1 where the PHT 60 $\mu$m image in grey scale is overlaid with the 6 cm radio continuum image in contours, the overall morphological resemblance between the infrared and radio images appears to be good out to at least the last contour at $r_0 \sim 5'$. The isophotal shapes are quite similar and intensity peaks outside the nucleus spatially coincide.

The radial profiles in infrared and radio continuum are compared in Fig. 2. Each data point represents a 40″ × 40″ cell as defined above. Open squares and crosses are data cells within 45° of the PHT in-scan and cross-scan directions (see Tuffs et al. 1996), respectively. For the PHT data, those filled circles are cells with $r_0 < 1'.5$ whose 60 $\mu$m SB were derived from the IRAS HiRes data. The separation of the "in-scan" cells from the "cross-scan" cells allows us for a check on the detector transient effect which, if present, would affect the in-scan cells much more severely than the cross-scan cells. We use the radio profiles as control samples for this purpose. The mean difference between the PHT in-scan and cross-scan profiles reaches up to $\sim$ 0.21 in log scale at $r \sim 5'$ with the in-scan consistently brighter than the cross-scan at the same radius. Less than half of this difference might also be observed in the radio profiles. Thus, the memory effect could overestimate the 60 $\mu$m SB at outer radii by $\sim$ 26% in the PHT in-scan direction. In the analysis below, we mainly rely on the PHT cross-scan data to make our best estimates on disk properties.

The galaxy disk at 60 $\mu$m appears also to follow an exponential law on average. This exponential form apparently holds even beyond the 25th surface magnitude in blue (Tuffs et al. 1996). Table 2 compares the e-folding disk scale-lengths, $r_s$, among the 60 $\mu$m, the total radio continuum [columns (3) and (4)] and the non-thermal radio component [columns (5) and (6)]. We determined $r_s$ and its $1\sigma$ statistical error (given in parentheses in Table 2) from a least-squares fit to the corresponding mean radial profile over $1'.5 < r_0 < 4'$.

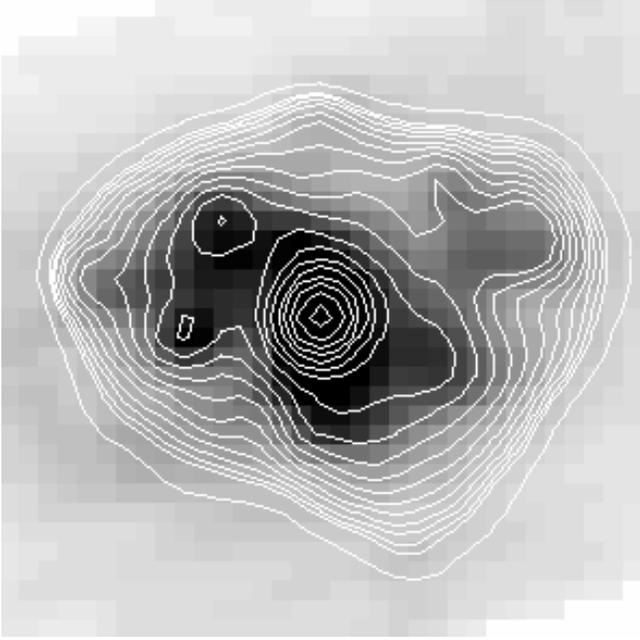

**Fig. 1.** PHT image of NGC 6946 at 60 $\mu$m in grey scale, overlaid with contours of its radio continuum image at 6 cm. Brightnesses are displayed on a linear scale. Both images have a size of $11' \times 11'$, an angular resolution of about $52''$, and a pixel size of $20'' \times 20''$. North is at top and East to the left.

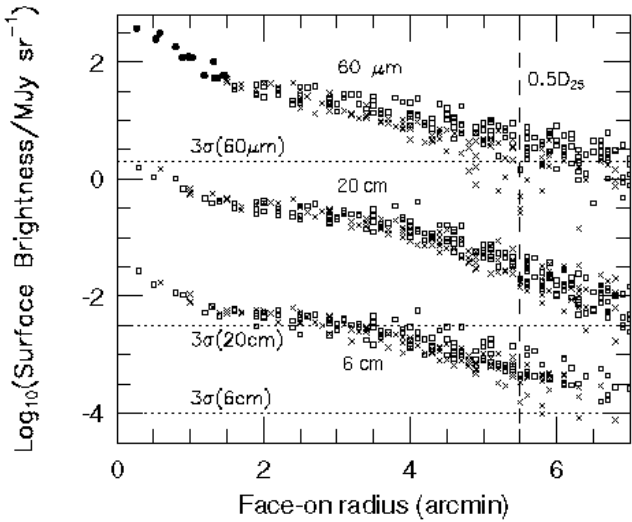

**Fig. 2.** Plot of logarithmic SB as a function of the face-on radius, $r_0$, for the 60 $\mu$m data, and for radio continuum data at 20 cm and 6 cm. The 6 cm radio continuum data points have been vertically shifted by $-1.5$ in log scale to be visually separated from the 20 cm data points. Each data point represents the mean SB within a cell of $40'' \times 40''$, with open squares and crosses for cells within $45°$ of the PHT in-scan and cross-scan directions, respectively. At 60 $\mu$m, the SB in cells with a radius less than $1'.5$ are derived from the IRAS data, and are plotted as filled circles. The dotted vertical line indicates $r_0 = 5'.5$ ($\approx D_{25}/2$) and dashed horizontal lines schematically indicate the $3\sigma$ levels.

The resulting $r_s$ from fitting to the cross-scan data cells [row (2) of Table 2] is always smaller than that from fitting to all the data cells [row (1)]. This is largely because of the relative concentration of HII regions near the in-scan direction (at P.A. of $107°$) over our fitting radii (see Fig. 1). For the PHT data, a contribution from the detector memory effect might also contribute to this $r_s$ difference. Because HII regions fall off at large radii, the derived scale-length would be smaller if one has fit the disk over a larger radial range (e.g., van der Kruit et al. 1977; Engargiola 1991). Nevertheless, Table 2 clearly shows that the disk scale-length at 60 $\mu$m is shorter than that in radio continuum at a significance level of $\gtrsim 2\sigma$, and that this is not simply because of a varying contribution from the thermal radio component. Similar variations of disk scale-ength with wavelength have also been found in other nearby galaxies using IRAS data (e.g., Rice et al. 1990).

**Table 2.** Disk exponential scale lengths in arc-minutes

| Data cells (1) | 60 $\mu$m (2) | 6 cm$^{tot}$ (3) | 20 cm$^{tot}$ (4) | 6 cm$^{nt}$ (5) | 20 cm$^{nt}$ (6) |
|---|---|---|---|---|---|
| All cells | 1.91 (0.16) | 2.33 (0.25) | 2.63 (0.31) | 3.01 (0.41) | 3.01 (0.41) |
| Cross-scan | 1.30 (0.11) | 1.81 (0.23) | 2.02 (0.28) | 2.26 (0.35) | 2.26 (0.35) |

## 4. Spatial Variation of the FIR-to-Radio Ratio

We define $Q_{20cm}$ ($Q_{6cm}$) to be the ratio of the SB at 60 $\mu$m to the total SB at 20 cm (6 cm). The spatial variation of $Q$ is illustrated in Fig. 3 as a function of $r_0$, and in Fig. 4 by plotting $Q_{20cm}$ as a function of the 6 cm SB. Only the cross-scan data points with $1'.5 < r_0 < 4'$ are shown here.

Some numerical results can be drawn from Figs. 3 and 4: (1) From $1'.5$ to $4'$ in radius, the PHT data show that $Q_{20cm}$ and $Q_{6cm}$ decrease by a factor of $\sim 1.7$ ($\pm 0.04$) and $\sim 1.5$ ($\pm 0.03$), respectively. The slower decreasing rate of $Q_{6cm}$ is consistent with the fact that over the radial range considered here, the fractional thermal radio component slowly decreases radially out. For the same reason, a removal of the thermal component from the radio continuum would make the radial profile of $Q$ even steeper. (2) Fig. 4 shows that $Q$ is also correlated with SB on average. As the 6 cm SB decreases, the same data as used in Fig. 3 shows an average drop by a factor of $1.9^{+0.27}_{-0.23}$ in Fig. 4. These results are in quantitative agreement with the claims by

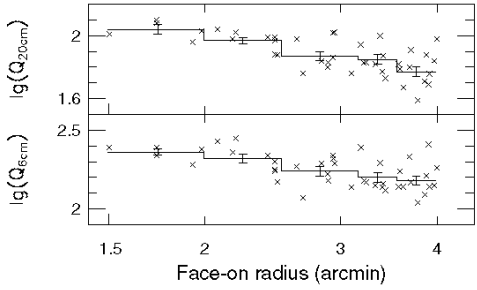

**Fig. 3.** Plots of the far-infrared to radio continuum ratio, Q, as a function of radius $r_0$. The upper panel shows 60 $\mu$m to 20 cm, while the lower panel shows 60 $\mu$m to 6 cm. Only the cross-scan data points with $1'.5 < r_0 < 4'$ are shown here. The solid line in each panel indicates the resulting mean radial profile of $Q$. Also shown are $1\sigma$ error bars for the means, estimated from the scatter of the data points in the figure.

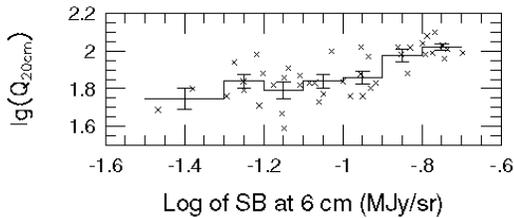

**Fig. 4.** Plot of $Q_{20cm}$ as a function of the SB at 6 cm using the same data set as used in Fig. 3. The solid line is the mean variation of $Q_{20cm}$ as the 6 cm SB changes, superposed with $1\sigma$ error bars for the means as defined in Fig. 3.

Marsh & Helou (1995), but are significantly inconsistent with that of Fitt et al. (1992).

We can show that, for NGC 6946, the thermal radio component is too small for the observed variation of $Q$ to be accounted for by a radially decreasing "contamination" to a "FIR-cold", diffuse component with a constant FIR-to-nonthermal radio continuum ratio of $Q_c$ from a "FIR-warm" component (e.g., HII regions) with a constant FIR-to-thermal radio continuum ratio of $Q_w$. Such a "two-component" model was first proposed by Xu et al. (1994), but for the global IR/radio correlation among disk galaxies. Under this model, one can write

$$Q_{20cm} = Q_w f_t + Q_c (1 - f_t), \qquad (1)$$

where $f_t$ is the fractional thermal radio component at 20 cm. For NGC 6946, $f_t$ changes on average from $\sim 18\%$ at $r_0 = 1'.5$ to $\sim 8\%$ at $r_0 = 4'$; and over the same radii, $Q_{20cm}$ drops by a factor of 1.7 to a value of 60 at $r_0 = 4'$ (see Fig. 3). By averaging eq. (1) at each of these radii and then taking the difference between them, one has

$$Q_w - Q_c \approx 415. \qquad (2)$$

On the other hand, $Q_{20cm}$ reaches a maximum of $\sim 125$ when $f_t$ approaches its maximum of $\sim 0.5$ at 20 cm. For these positions, eq. (1) gives

$$Q_w + Q_c \approx 250. \qquad (3)$$

Obviously, eqs. (2) and (3) together have no non-negative solution for $Q_c$.

By contrast, our results are qualitatively in agreement with the description of the radio disk as a "smeared" version of the infrared disk of a galaxy (Bicay & Helou 1990). Each local maximum in disk brightness adds dispersion to the plot of $Q$ vs. radius, but is a local center of star formation from which $Q$ decreases radially. The radio disk arises primarily from synchrotron emission by cosmic ray electrons in the galaxy's magnetic field, whereas dust heated by stars defines the infrared disk. Cosmic ray electrons can diffuse away from their sources as much as a kpc before decaying radiatively; heating photons on the other hand will typically get absorbed and re-emitted in the infrared within a hundred pc of their source. Assuming that both radiating electrons and heating photons originate in the same sites of star formation, this difference in radiating scale length explains the "smearing" relation between infrared and radio disks (Helou & Bicay 1993).

*Acknowledgements.* We wish to thank R. Beck for providing us with the radio images and the referee for a number of useful comments. This work was supported in part by ISO data analysis funding from the US National Aeronautics and Space Administration, and carried out at the Infrared Processing and Analysis Center and the Jet Propulsion Laboratory of the California Institute of Technology.